\documentclass[
preprintnumbers, 
showpacs, 
amsmath, 
showkeys, 
amssymb, 
aps, 
superscriptaddress, 
longbibliography, 
letterpaper,
twocolumn, 
]{revtex4-2}
\usepackage{lipsum, babel}
\usepackage{color}
\usepackage{hyperref}
\usepackage{bm}
\usepackage{amsfonts}
\usepackage{mathrsfs}
\usepackage{graphicx}
\usepackage{amsmath}
\usepackage{float}
\usepackage{braket}
\usepackage{natbib}

     \hypersetup{
    colorlinks=true,
    linkcolor=red,
    citecolor=blue,
}

\begin{document}

\newcommand{\sgn}{\operatorname{sgn}}
\newcommand{\hhat}[1]{\hat {\hat{#1}}}
\newcommand{\pslash}[1]{#1\llap{\sl/}}
\newcommand{\kslash}[1]{\rlap{\sl/}#1}
\newcommand{\lab}[1]{}
\newcommand{\sto}[1]{\begin{center} \textit{#1} \end{center}}
\newcommand{\rf}[1]{{\color{blue}[\textit{#1}]}}
\newcommand{\eml}[1]{#1}
\newcommand{\el}[1]{\label{#1}}
\newcommand{\er}[1]{Eq.\eqref{#1}}
\newcommand{\df}[1]{\textbf{#1}}
\newcommand{\mdf}[1]{\pmb{#1}}
\newcommand{\ft}[1]{\footnote{#1}}
\newcommand{\n}[1]{$#1$}
\newcommand{\fals}[1]{$^\times$ #1}
\newcommand{\new}{{\color{red}$^{NEW}$ }}
\newcommand{\ci}[1]{}
\newcommand{\de}[1]{{\color{green}\underline{#1}}}
\newcommand{\ke}{\rangle}
\newcommand{\br}{\langle}
\newcommand{\lb}{\left(}
\newcommand{\rb}{\right)}
\newcommand{\lbk}{\left[}
\newcommand{\rbk}{\right]}
\newcommand{\blb}{\Big(}
\newcommand{\brb}{\Big)}
\newcommand{\nn}{\nonumber \\}
\newcommand{\p}{\partial}
\newcommand{\pd}[1]{\frac {\partial} {\partial #1}}
\newcommand{\cd}{\nabla}
\newcommand{\cc}{$>$}
\newcommand{\bqa}{\begin{eqnarray}}
\newcommand{\eqa}{\end{eqnarray}}
\newcommand{\bqe}{\begin{equation}}
\newcommand{\eqe}{\end{equation}}
\newcommand{\bay}[1]{\left(\begin{array}{#1}}
\newcommand{\eay}{\end{array}\right)}
\newcommand{\eg}{\textit{e.g.} }
\newcommand{\ie}{\textit{i.e.}, }
\newcommand{\iv}[1]{{#1}^{-1}}
\newcommand{\st}[1]{|#1\ke}
\newcommand{\at}[1]{{\Big|}_{#1}}
\newcommand{\zt}[1]{\texttt{#1}}
\newcommand{\non}{\nonumber}
\newcommand{\m}{\mu}
\def\xa{{m}}
\def\xA{{m}}
\def\xb{{\beta}}
\def\xB{{\Beta}}
\def\xd{{\delta}}
\def\xD{{\Delta}}
\def\xe{{\epsilon}}
\def\xE{{\Epsilon}}
\def\xve{{\varepsilon}}
\def\xg{{\gamma}}
\def\xG{{\Gamma}}
\def\xk{{\kappa}}
\def\xK{{\Kappa}}
\def\xl{{\lambda}}
\def\xL{{\Lambda}}
\def\xo{{\omega}}
\def\xO{{\Omega}}
\def\xvp{{\varphi}}
\def\xs{{\sigma}}
\def\xS{{\Sigma}}
\def\xt{{\theta}}
\def\xvt{{\vartheta}}
\def\xT{{\Theta}}
\def \Tr {{\rm Tr}}
\def\CA{{\cal A}}
\def\CC{{\cal C}}
\def\CD{{\cal D}}
\def\CE{{\cal E}}
\def\CF{{\cal F}}
\def\CH{{\cal H}}
\def\CJ{{\cal J}}
\def\CK{{\cal K}}
\def\CL{{\cal L}}
\def\CM{{\cal M}}
\def\CN{{\cal N}}
\def\CO{{\cal O}}
\def\CP{{\cal P}}
\def\CQ{{\cal Q}}
\def\CR{{\cal R}}
\def\CS{{\cal S}}
\def\CT{{\cal T}}
\def\CV{{\cal V}}
\def\CW{{\cal W}}
\def\CY{{\cal Y}}
\def\BC{\mathbb{C}}
\def\BR{\mathbb{R}}
\def\BZ{\mathbb{Z}}
\def\sA{\mathscr{A}}
\def\sB{\mathscr{B}}
\def\sF{\mathscr{F}}
\def\sG{\mathscr{G}}
\def\sH{\mathscr{H}}
\def\sJ{\mathscr{J}}
\def\sL{\mathscr{L}}
\def\sM{\mathscr{M}}
\def\sN{\mathscr{N}}
\def\sO{\mathscr{O}}
\def\sP{\mathscr{P}}
\def\sR{\mathscr{R}}
\def\sQ{\mathscr{Q}}
\def\sS{\mathscr{S}}
\def\sX{\mathscr{X}}

\author{Wei Kou}
\email{kouwei@impcas.ac.cn}
\affiliation{Institute of Modern Physics, Chinese Academy of Sciences, Lanzhou 730000, Gansu Province, China}
\affiliation{School of Nuclear Science and Technology, University of Chinese Academy of Sciences, Beijing 100049, China}
\author{Bing'ang Guo}
\email{guobingang@impcas.ac.cn}
\affiliation{Institute of Modern Physics, Chinese Academy of Sciences, Lanzhou 730000, Gansu Province, China}
\affiliation{School of Nuclear Science and Technology, University of Chinese Academy of Sciences, Beijing 100049, China}
\author{Xurong Chen}
\email{xchen@impcas.ac.cn (Corresponding Author)}
\affiliation{Institute of Modern Physics, Chinese Academy of Sciences, Lanzhou 730000, Gansu Province, China}
\affiliation{School of Nuclear Science and Technology, University of Chinese Academy of Sciences, Beijing 100049, China}
\affiliation{Southern Center for Nuclear Science Theory (SCNT), Institute of Modern Physics, Chinese Academy of Sciences, Huizhou 516000, Guangdong Province, China}

\title{Unveiling the Secrets of Vortex Neutron Decay}

\begin{abstract}

Investigation of decay and scattering processes of particles in a vortex state offers a novel and promising approach for probing particle structure. Our study reveals distinct properties of vortex neutron decay, which deviate from those of classical plane-wave neutron decay. We present the energy-angle distributions of the final-state electron and antineutrino in unpolarized vortex neutrons, as well as angle distributions integrated over their energies. Notably, we provide theoretical calculations of the decay behavior of neutrons with varying vortex cone angles and initial energies. We propose that identifying the vortex state of the initial neutron can be achieved by analyzing the angular and energy distributions of the final-state particles, introducing new degrees of freedom for studying weak interactions and neutron decay kinematics that have been previously overlooked in particle physics. This timely investigation takes advantage of recent advancements in vortex neutron preparation and analysis, opening up new avenues for exploring the fundamental properties of matter.
\end{abstract}


\maketitle

\section{Introduction}
\label{sec:introduction}
Research on the beta decay of free neutrons reveals the characteristics of weak interactions in the field of fundamental particle physics. The beta decay refers to a neutron decaying into a proton, while emitting an electron and the corresponding electron antineutrino. Many experiments associated with beta decay provide empirical evidence for exploring the electroweak interactions \cite{Glashow:1961tr,Weinberg:1967tq,Salam:1968rm} in the Standard Model. Additionally, the theoretical calculations of neutron beta decay at the tree level are also considered a classic model for studying the theory of weak interactions \cite{Fermi:1932xva,Fermi:1934hr}. 

In traditional high-energy particle physics research, the focus has mainly been on plane-wave particles, where the particle wave function (or field) does not contain intrinsic orbital angular momentum (OAM) information. In recent years, discussions in optics, electromagnetism, atomic physics, and condensed matter physics have explored the differences in interactions when particles with non-zero intrinsic OAM are involved compared to the plane-wave case. Vortex states, also known as twisted states of fields or wave functions, are non-plane-wave solutions of the corresponding wave equation that propagate along a fixed direction (axis $z$) with an average non-zero $z$-direction projection of intrinsic OAM. The OAM mentioned here is not the relative OAM between composite particles, nor is it the OAM generated by the motion of particles in an external field. The OAM of vortex state fields is entirely an intrinsic property of the free field. The vortex wave is a monochromatic wave with helicoidal, corkscrew-like features, and the wavefront is generated by the azimuthal phase factor $\exp(i\ell\varphi_r)$, where $\ell$ is a nonzero OAM. Detailed information on vortex states can be found in Ref. \cite{Ivanov:2022jzh}. 

Over the past few decades, vortex photons have been well understood and extensively studied for various applications \cite{Allen:1992zz,Molina-Terriza:2007ydx,zhan2009cylindrical,torres2011twisted,harris2015structured,padgett2017orbital}. In recent years, vortex electrons have also been experimentally prepared \cite{Uchida:2010hbm,Verbeeck:2010ezk,McMorran:2011bql}. These vortex states are widely applied in atomic and materials physics \cite{Bliokh:2017uvr,lloyd2017electron}. However, the introduction of vortex states in high-energy particle physics has only gained attention in the past decade (see Ref. \cite{Ivanov:2022jzh} and references therein). The studies on vortex state particle collisions \cite{Ivanov:2019pdt} and decays \cite{Zhao:2021joa} are provided here for further reading. In high-energy physics, the focus has traditionally been on energy, momentum, transverse momentum distribution, and spin distribution of particles. The addition of intrinsic OAM as a new degree of freedom enables the exploration of previously unconsidered particle characteristics in high-energy vortex particle scattering and decay experiments. By measuring the final-state particles, new properties introduced by this degree of freedom can be identified.

The recent preparation of neutral particle vortex states, such as neutrons \cite{Clark:2015rcq,sarenac2018methods,sarenac2019generation} and heavier composite particles like atoms \cite{luski2021vortex}, has generated significant interest, opening new applications in particle physics. Although experimentally imparting vortex information to heavier particles is challenging, theoretical advancements continue. Vortex neutrons, with their intrinsic OAM projection along the propagation direction, offer new degrees of freedom compared to classical neutrons. Their properties—electrical neutrality, magnetic sensitivity, and strong penetrating power—make them valuable for detecting material electromagnetic properties. Vortex neutrons are crucial for probing the internal structure of neutrons, especially their electromagnetic structure \cite{larocque2018twisting}. They also play key roles in nuclear processes \cite{Afanasev:2019rlo} and nucleon scattering \cite{Afanasev:2021uth}, enabling the detection of phase differences between nucleon participation in strong and electromagnetic interactions. Furthermore, discussions also cover the potential of vortex neutrons as material probes and other related topics \cite{Thien:2022coj,sarenac2024smallangle}.

This study employs vortex states of neutrons to investigate the neutron decay process. The beta decay spectrum of conventional neutrons and related corrections are well-documented in textbook discussions \cite{griffiths2020introduction}. Building upon calculations of plane wave neutron decay, we incorporate vortex state information to present a novel spectrum for vortex neutron decay. For the first time, this work provides electron and neutrino energy spectra resulting from vortex state neutron decay and compares spectra differences for various neutron energies and vortex cone angles. We propose that it may be possible to experimentally differentiate whether the initial neutron is in a vortex state based on the beta decay spectrum. 
The structure of this article is as follows: In Sec. \ref{sec:2}, we briefly provide calculations for plane wave neutron beta decay, which sets the groundwork for extending the analysis to vortex state neutron decay. We then incorporate vortex information into beta decay and discuss the characteristics of non-polarized vortex neutron decay. In Sec. \ref{sec:4}, we present and discuss numerical calculation results. Finally, we provide conclusions and a future outlook.

\section{Decay of the plane-wave neutrons}
\label{sec:2}
\begin{figure}[htbp]
	\centering
	\includegraphics[width=0.3\textwidth]{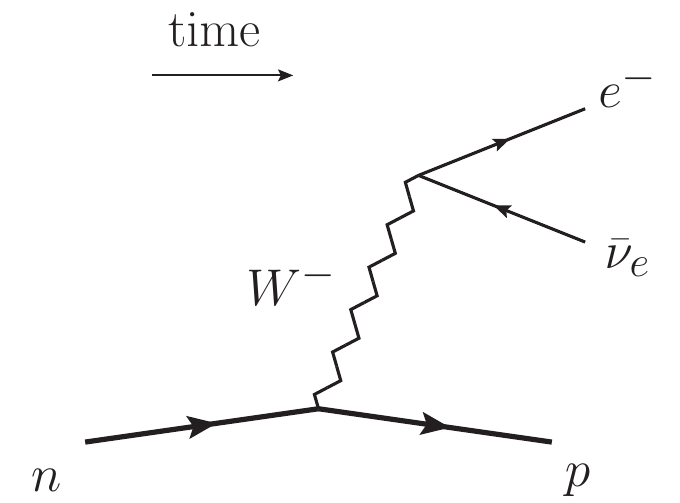}
	\caption{The Feynman diagram for neutron beta decay at the tree level. Neutrons and protons are assumed to be point-like particles, without considering their internal structure.}
	\label{fig:beta decay}
\end{figure}

Within the Standard Model framework, neutron beta decay calculations at the tree level rely on Fermi's four-fermion weak interaction theory \cite{Fermi:1932xva,Fermi:1934hr}. Introducing the $W^-$ boson into these calculations is also consistent with the Standard Model. Despite slight discrepancies between tree-level results and experiments, they remain central to this work. This is because the $V-A$ theory \cite{Sudarshan:1958vf} and CKM matrix elements \cite{PhysRevLett.10.531,Kobayashi:1973fv} related to quark flavor changes do not affect our description of vortex neutrons. The beta decay process of a free neutron is given by $n(p)\to e^-(k)+\bar{\nu}_e(q_1)+p(q_2)$ (see Figure \ref{fig:beta decay}).
Since most of the neutron's energy is used to produce a proton, the electron's energy is comparable to its mass, making it essential to account for the electron mass. The rest masses of the neutron, proton, and electron are $m_n=939.57$ MeV, $m_p=938.27$ MeV, and $m_e=0.511$ MeV, respectively, with $\hbar=c=1$. The four-momentum of the neutron is represented as $p^\mu=(E_n,|\vec{p}|\vec{n})=(E_n,E_n\beta_n\vec{n})$, where $|\vec{p}|=\sqrt{E_n^2-m_n^2}$ and $\beta_n=\sqrt{1-\frac{1}{\gamma_n^2}}$ is the neutron's speed with Lorentz factor $\gamma_n$. Similarly, the four-momentum of the final-state electron (the electron spectrum is measured, and the neutrino spectrum can be directly extended) is $k^\mu=(E_e,|\vec{k}|\vec{n}_e)$. The four-momenta of the neutrino and proton are denoted as $q_1^\mu$ and $q_2^\mu$, respectively, with the invariant mass square given by $q^2=(q_1+q_2)^2$. Evidently, the relation $p-k=q=q_1+q_2$ holds.

According to Feynman rules, the $\mathcal{S}$ matrix element for the neutron decay at tree-level as shown in Figure \ref{fig:beta decay} is given by:
\begin{equation}
	\begin{aligned}
		\mathcal{S}&=i\mathcal{M}(2\pi)^4\delta^4(p-k-q_1-q_2),\\
		\mathcal{M}&=\sqrt{\frac{1}{2}}G_F \bar{u}(q_2)O^\alpha u(p) \bar{u}(k)O_\alpha v(q_1),
	\end{aligned}
	\label{eq:matrix}
\end{equation}
Here, $G_F\equiv\frac{\sqrt{2}}{8}(g_W/m_W)^2\simeq 1.166\times 10^{-5}$ GeV$^{-2}$ represents the Fermi constant. The operator $O_\alpha=\gamma_\alpha(1+\gamma_5)$ is the chiral operator involving the Gamma matrices. To simplify the formalism without affecting the conclusions of this work, we do not include corrections related to the Cabibbo angle and the vector and axial vector weak charges, the extension of the scattering amplitude is evident. Summing over all spin states results in the square of the scattering amplitude:
\begin{equation}
	\langle|\mathcal{M}|^2\rangle=128G_F^2(p\cdot q_1)(k\cdot q_2).
	\label{eq:amplitude}
\end{equation}
This quantity is independent of Lorentz spacetime transformations. Consequently, the differential width of the three-body beta decay can be expressed as:
\begin{equation}
	\begin{aligned}
		d\Gamma=&\frac{\langle|\mathcal{M}|^2\rangle}{2\cdot 2E_n}(2\pi)^4\delta^4(p-k-q_1-q_2)\\
		&\times \frac{d^3\vec{k}}{(2\pi)^32E_e} \frac{d^3\vec{q}_1}{(2\pi)^32E_1} \frac{d^3\vec{q}_2}{(2\pi)^32E_2},
	\end{aligned}
	\label{eq:width}
\end{equation}
Here, $E_1$ and $E_2$ represent the energies of the antineutrino and proton, respectively. By integrating over the momenta of the final-state neutrino and proton, the neutron decay spectrum as a function of the final-state electron energy can be obtained. To derive the neutrino or proton spectrum, integration over the momenta of the final-state particles other than the corresponding particle is necessary.

When measuring the final-state electron, integrating over the momenta of the neutrino and proton results in a differential width that solely depends on the inner product of the initial neutron and final electron four-momenta, along with the energies of the neutron and electron. This quantity is Lorentz invariant, meaning its form remains unchanged regardless of the chosen reference frame. From calculation, we get
	\begin{equation}
		\begin{aligned}
			&\frac{d\Gamma_{PW}}{dE_ed\Omega}=\frac{G_F^2\sqrt{E_e^2-m_e^2}\left(m_n^2-m_p^2+m_e^2-2(p\cdot k)\right)^2}{48\pi^4 E_n\left(m_n^2+m_e^2-2(p\cdot k)\right)^3}\\
			&\quad\quad\times(A+B+C),\\
			&A=\left(3m_e^4+10m_e^2m_n^2+3m_n^4+3m_p^2(m_e^2+m_n^2)\right)(p\cdot k),\\
			&B=8(p\cdot k)^3-2\left(5(m_e^2+m_n^2)+m_p^2\right)(p\cdot k)^2,\\
			&C=-2m_e^2m_n^2\left(m_e^2+m_n^2+2m_p^2\right),
		\end{aligned}
		\label{eq:fullwidth}
	\end{equation}
	where $\Omega$ represents the electron scattering solid angle. 
	
	If the neutron is at rest, then $(p\cdot k)=m_nE_e$ and $E_n=m_n$. Substituting into Eq. (\ref{eq:fullwidth}), one can obtain the differential decay width for a stationary neutron, which is consistent with the results in Ref. \cite{griffiths2020introduction}. Additionally, if the neutron has initial kinetic energy or speed, the corresponding Lorentz factor is $\gamma_n=1/\sqrt{1-\beta_n^2}$, then $(p\cdot k)=E_n(E_e-\beta_n|\vec{k}|\vec{n}\cdot\vec{n}_e)$, where $E_n=\gamma_n m_n$ and $\vec{n}\cdot\vec{n}_e=\cos\theta$ with the emitted electron scattering angle $\theta$. One can verify that when the initial neutron has a Lorentz factor $\gamma_n$, the total decay width (integrating over electron energy and solid angle) is related to the total width of the decay of a stationary neutron as $\Gamma_{PW}=\Gamma_0/\gamma_n$, where $\Gamma_0$ is the total width of beta decay of a stationary neutron. This is a manifestation of the relativistic length contraction effect.

\section{Unpolarized vortex neutron case}
\label{sec:3}
\subsection{Constructing the unpolarized vortex neutron}
Compared to plane-wave neutron states, vortex neutrons have new degrees of freedom, namely the intrinsic OAM information. We use the simplest vortex state - the Bessel vortex state as an example to describe the decay of unpolarized vortex neutron states. This is because in high-energy physics, particles prepared in Bessel vortex states have definite energy, longitudinal momentum, transverse momentum modulus, and a non-zero OAM. These are important properties in the kinematics of particles in high-energy physics research. Moreover, the Bessel vortex states of leptons, photons, and some hadrons have been applied in the theoretical analysis of high-energy physics scattering and decay processes.

\begin{figure}[h]
	\centering
	\includegraphics[width=0.45\textwidth]{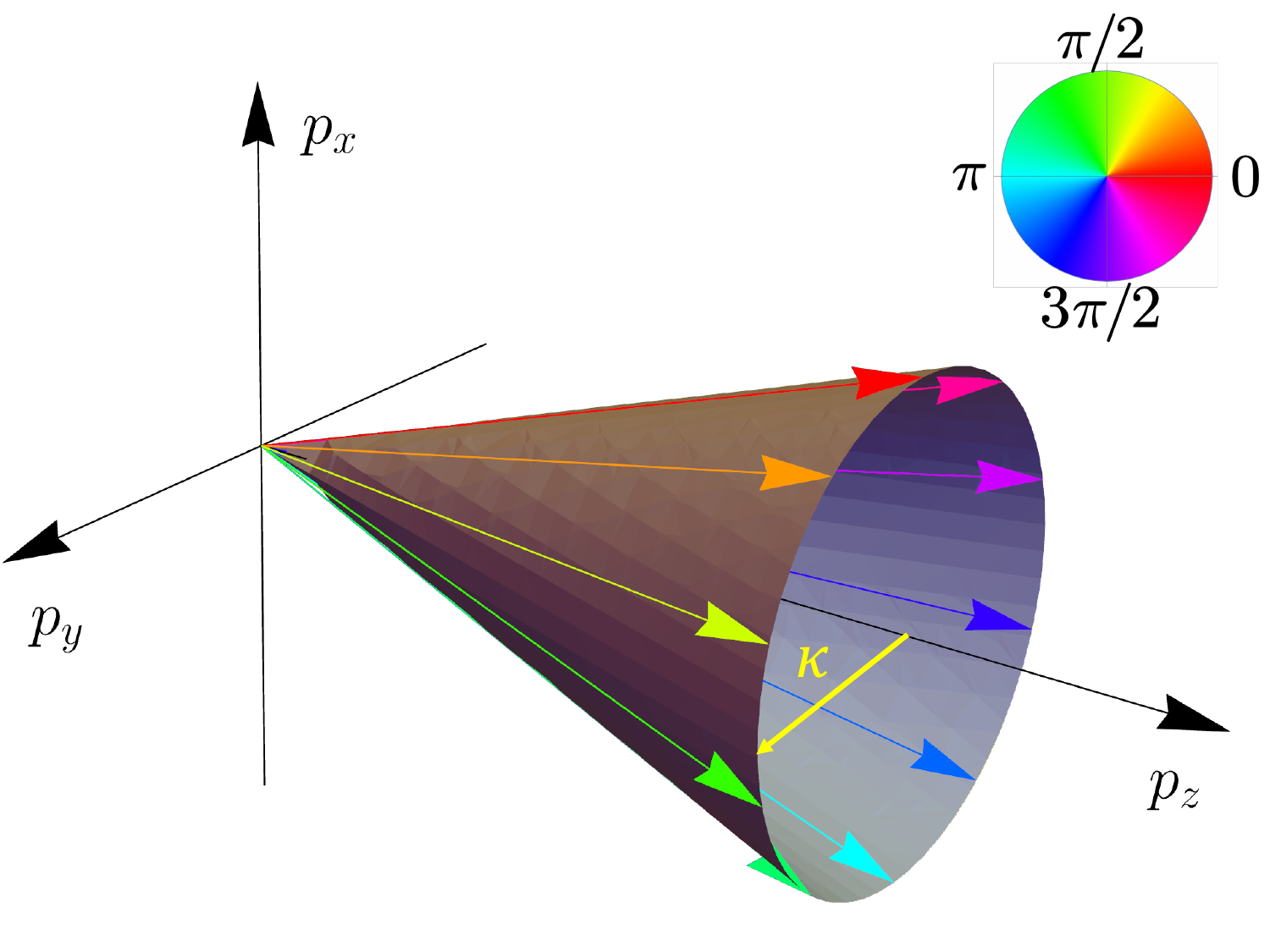}
	\caption{The distribution of Bessel vortex states in momentum space. The ring radius $\kappa$ represents the transverse momentum modulus of the vortex state, and the conical surface is composed of the momenta of all plane-wave components of the vortex state, with different colors representing different vortex phase factors.}
	\label{fig:cone}
\end{figure}

The wave function of the Bessel vortex state is an exact solution of the corresponding Schrödinger equation and Klein-Gordon equation, and the Bessel vortex states with different quantum numbers form a complete basis for the solution space of the wave equation. This indicates that the wave function of the Bessel vortex state belongs to the monochromatic solution of the wave equation and is composed of a superposition of plane waves with the same energy $E$, longitudinal momentum $p_z$, and the same transverse momentum modulus $|\vec{p}_{T}|=\kappa$, but with different azimuthal angles $\varphi_p$. Figure \ref{fig:cone} provides a visual representation of these characteristics, and specific definitions of the Bessel vortex states can be found in the relevant review \cite{Ivanov:2022jzh}.

For neutrons in Bessel vortex states, the momenta corresponding to all plane-wave components rotate around an axis and are evenly distributed on a conical surface. The opening angle of the cone, $\theta_0$, is related to the projections of the momentum in the transverse and longitudinal directions, i.e.,
\begin{equation}
	\cos\theta_0\equiv\frac{p_z}{|\vec{p}|},\qquad \sin\theta_0\equiv\frac{\kappa}{|\vec{p}|}.
\end{equation}
Ideally, of course, one would expect this opening angle to be larger, as it can bring more information different from that of plane-wave states. However, in terms of experimental implementation, this cone angle is very small, typically only a few degrees.

The construction of fermionic vortex states is based on scalar particle vortex states, which can be simply understood as the product of a portion of plane waves with the phase factor corresponding to the vortex azimuthal angle. This is interpreted as the distribution of all plane wave momenta forming a conical surface, which can be expressed as \cite{Jentschura:2010ap,Jentschura:2011ih}
\begin{equation}
	\left|\Psi_{E,p_z,\ell}^B(\boldsymbol{x},t)\right\rangle\propto J_\ell(\kappa r)\mathrm{exp}(i\ell\varphi+ip_zz-iEt),
	\label{eq:Bessel}
\end{equation}
where $J_\ell$ is the Bessel functions of the first kind. This form is a result in coordinate space, and it can be seen that the exponential part can be divided into a product of plane wave components and vortex phase factors. We transform it into momentum space and simplify to obtain as
\begin{equation}
	|\kappa,\ell\rangle\propto\int d\varphi_p \left|\Psi_{E,p_z}^B(\boldsymbol{p})\right\rangle \exp(i\ell\varphi_p).
\end{equation}
This is a simple form, representing the dependence of the vortex state in momentum space on the plane waves in momentum space $\left|\Psi_{E,p_z}^B(\boldsymbol{p})\right\rangle$ and the vortex azimuthal angle $\varphi_p$ corresponding to the momentum $\boldsymbol{p}$. 

The fermion vortex state is introduced by incorporating spinor wave functions into the scalar particle vortex state. However, the OAM operator and the spin operator do not commute with the Hamiltonian separately. Nevertheless, the total angular momentum operator formed by the sum of OAM and spin operators commutes with the Hamiltonian, so the quantum numbers of total angular momentum are generally used to define fermion vortex states \cite{Serbo:2015kia,Bialynicki-Birula:2016unl}. Additionally, with the introduction of spin, polarized vortex Dirac fields can be defined for different polarization scenarios. This is not the focus of this work; due to the rotational symmetry of the vortex neutron decay initial state under consideration, the kinematic analysis of the entire decay process has been simplified. This simplification is mainly reflected in the absence of interference when there is only one vortex state particle in the initial state, causing the phase factor related to angular momentum $\exp(i\ell\varphi)$ to disappear when calculating the amplitude modulus squared. Consequently, the decay process of the vortex state does not depend on the intrinsic OAM of the vortex state. However, due to the unique momentum distribution of the vortex state, it is still feasible to distinguish different kinematic features in the final state of neutron beta decay.

\subsection{Neutron beta decay with vortex state}
\label{subsec: energy}
As mentioned earlier, each plane wave component of the vortex neutron decays to the final state with its own kinematic structure, and interference phenomena do not occur. Therefore, one can directly apply the formalism of plane wave beta decay for moving neutrons. This leads to the differential width of Bessel vortex neutron decay as the average of the sum of azimuthal angle information corresponding to all plane wave components \cite{Ivanov:2011kk}, i.e.,
\begin{equation}
	d\Gamma_V=\int\frac{d\varphi_{{p}}}{2\pi}d\Gamma_{PW}(\vec{p}).
	\label{eq:inte phip}
\end{equation}
The decay width of the plane wave in the above equation can be obtained from the calculation process in Sec. \ref{sec:2}. However, due to the introduction of Bessel vortex states, the momentum of the initial neutron is not directed towards the direction of plane wave propagation, but instead has a momentum azimuthal angle distribution with a cone opening angle of $\theta_0$. In this case, if we consider the direction of electron emission as $\vec{n}_e=(\sin\theta,0,\cos\theta)$, the momentum direction of the initial neutron changes from $\vec{n}=(0,0,1)$ to $\vec{n}_p=(\sin\theta_0\cos\varphi_p,\sin\theta_0\sin\varphi_p,\cos\theta_0)$. In other words, the angle between the initial neutron and the outgoing electron changes from $\vec{n}\cdot\vec{n}_e=\cos\theta$ to $\vec{n}_p\cdot\vec{n}_e=\sin\theta\sin\theta_0\cos\varphi_p+\cos\theta\cos\theta_0$.

We now analyze the kinematic quantities of the neutron beta decay process in plane wave, namely the energies of the final-state electron and antineutrino. Throughout the process, momentum conservation is inevitable, leading to the constant relation $p-k=q=q_1+q_2$. Hence, there is the relationship $q^2=(p-k)^2=m_n^2+m_e^2-2(p\cdot k)$. If one considers a fixed scattering angle for the outgoing electron, it is evident that the differential decay width does not exist in the region where the electron energy is less than the electron's rest mass, and the maximum value of the electron energy depends on the initial energy of the plane wave neutron. Furthermore, by considering the vortex neutron, the maximum electron energy is further related to the cone opening angle of the vortex state. To maximize the energy of the final-state electron, it is necessary to ensure that the square of the four-momentum difference between the neutron and the electron in the momentum distribution of the three final particles is exactly the square of the mass threshold for producing a proton and a neutrino, i.e., $(p-k)^2\geq m_p^2$. Substituting the four-momentum components of the neutron and electron, one can obtain that in the case of plane wave neutron decay, the maximum energy of the final-state electron is
\begin{widetext}
	\begin{equation}
		E_{e,\mathrm{max},PW}=\frac{m_{n}^{2}+m_{e}^{2}-m_{p}^{2}+\beta_n\cos\theta\sqrt{(m_{e}^{2}+m_{n}^{2}-m_{p}^{2})^{2}-4E_n^{2}m_{e}^{2}(1-\beta_n^{2}\cos^{2}\theta)}}{2E_n(1-\beta_n^{2}\cos^{2}\theta)}.
		\label{eq:Emaxpw}
	\end{equation}

Similarly, considering the final-state neutrino spectrum, one can obtain that the maximum energy of the neutrino can reach 
\end{widetext}
\begin{equation}
	E_{\nu,\max,PW}=\frac{m^2-(m_p+m_e)^2}{2E_n(1-\beta_n\cos\theta_\nu)}.
	\label{eq:Emaxpwnu}
\end{equation}

Let us consider the case of a vortex initial neutron and examine how it affects the maximum energy of the final-state electron or antineutrino. In contrast to the situation of plane wave neutron decay where the outgoing electron energy has only one threshold ($E_{e,\mathrm{max},PW}$), the energy threshold for electrons determined by the direction of decay of the vortex neutron can be defined in two scenarios
\begin{equation}
	\begin{aligned}
		E_{\mathrm{max},1}&=E_{e,\mathrm{max},PW}(\theta\to \theta+\theta_0),\\
		E_{\mathrm{max},2}&=E_{e,\mathrm{max},PW}(\theta\to \theta-\theta_0).
	\end{aligned}
	\label{eq:E1,E2}
\end{equation}
This means replacing the angle $\theta$ in Eq. (\ref{eq:Emaxpw}) with $\theta+\theta_0$ and $\theta-\theta_0$, respectively. This indicates that the two energy thresholds depend on the electron emission angles and are closely related to the cone angle and energy of the initial vortex state. These two electron energy thresholds divide the electron energy range of the final state of the vortex state decay into two cases \cite{Zhao:2021joa}, (i) when $m_e\leq E_e\leq 	E_{\mathrm{max},1}$, all plane wave components of the vortex state contribute to the final-state electron in this energy range (with $\theta$ fixed), and the azimuthal angle of the vortex ranges from $-\pi$ to $\pi$; (ii) when $E_{\mathrm{max},1}\leq E_e\leq 	E_{\mathrm{max},2}$, only some plane wave components of the vortex state contribute to the final-state electron in this energy range (with $\theta$ fixed), and the azimuthal angle of the vortex ranges from $-\omega$ to $\omega$, where the parameter $\omega$ is determined by the electron energy and the two threshold energies
\begin{equation}
	\cos\omega=\frac{(E_e-E_{\mathrm{max},1})E_{\mathrm{max},2}-(E_{\mathrm{max},2}-E_e)E_{\mathrm{max},1}}{E_e(E_{\mathrm{max},2}-E_{\mathrm{max},1})}
\end{equation}
with $0<\omega\leq \pi$. The situation is similar for the electron antineutrino.

\begin{figure*}[htbp]
	\centering
	\includegraphics[width=0.45\textwidth]{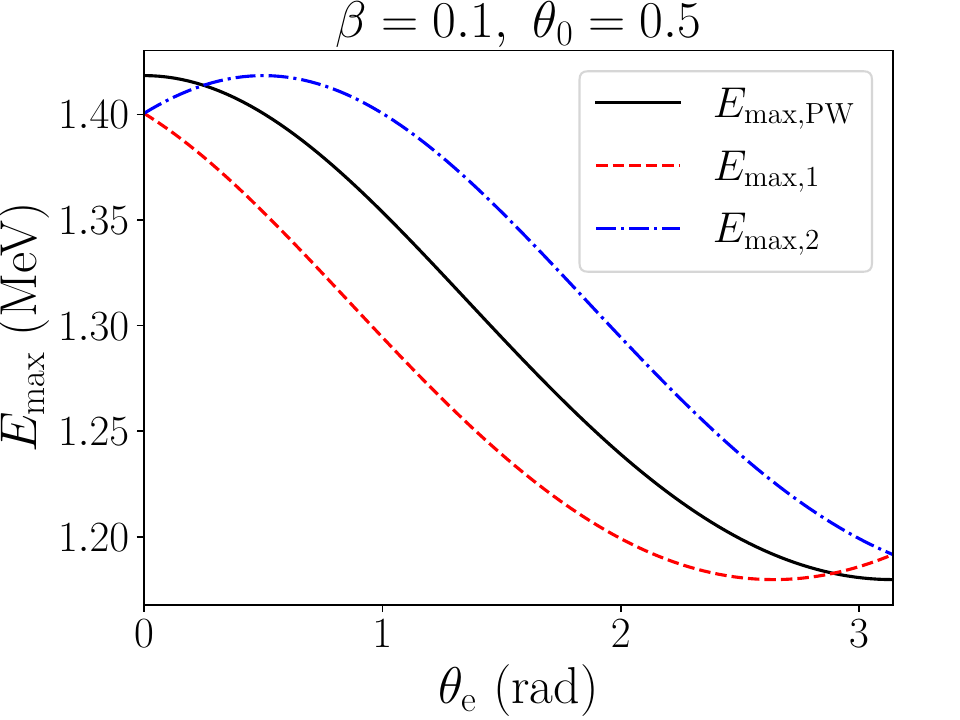}
	\includegraphics[width=0.45\textwidth]{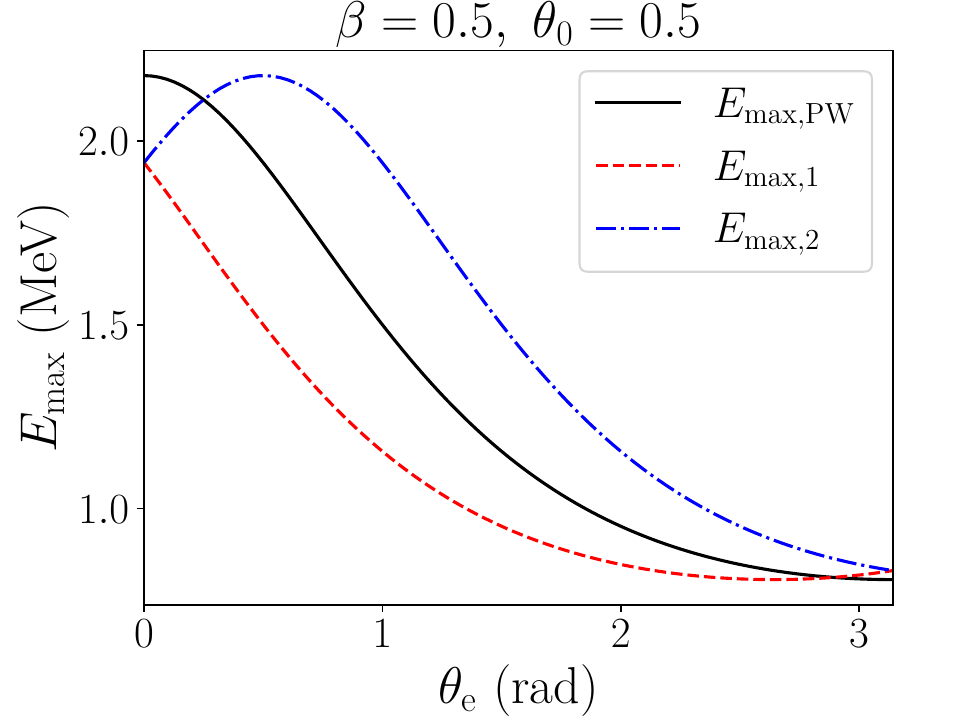}
	\includegraphics[width=0.45\textwidth]{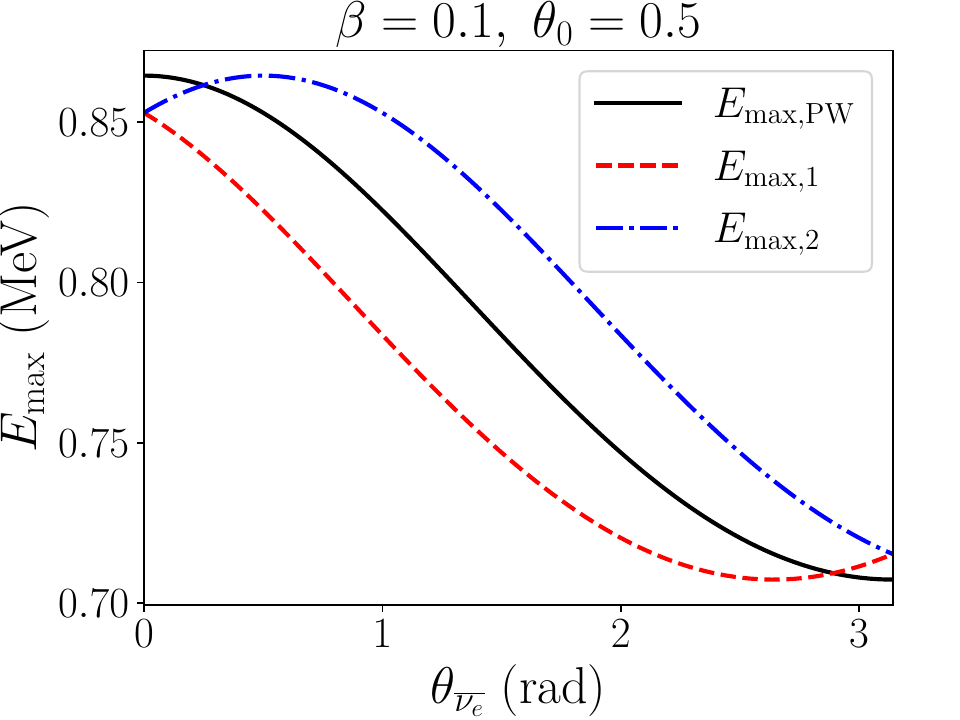}
	\includegraphics[width=0.45\textwidth]{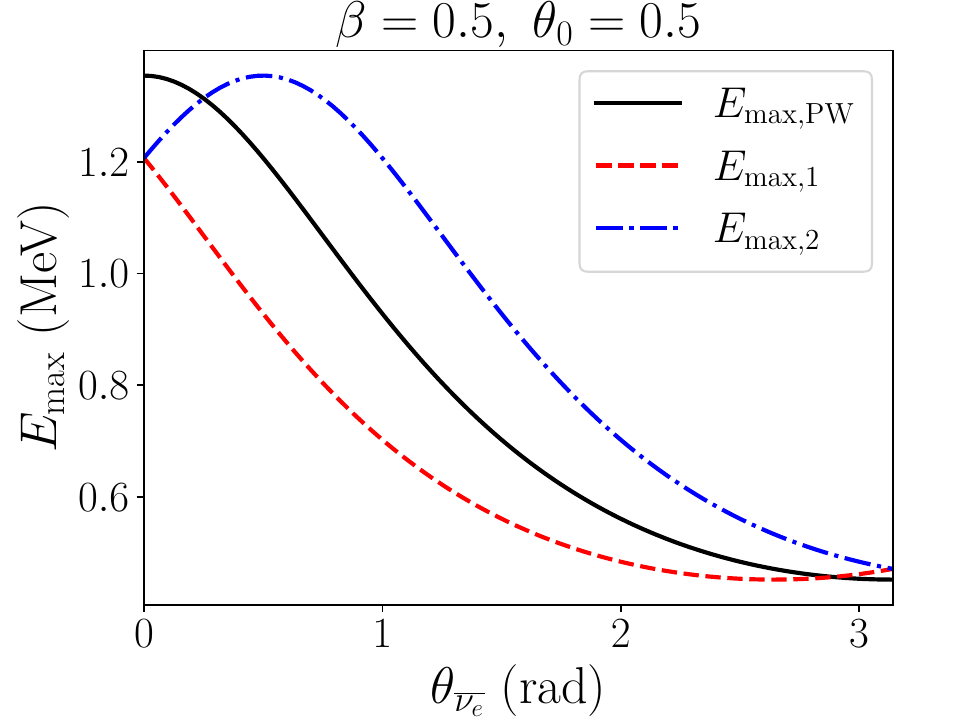}
	\caption{The energy ranges of the final-state electron and neutrino in beta decay are functions of their scattering angles, with the vortex cone opening angle fixed at $\theta_0=0.5$ rad. Assuming neutron speeds of $\beta_n=0.1$ and $0.5$.  (Top panel) Electron energy range. The black solid line represents the case of a plane wave neutron, while the red dashed line and blue dashed-dotted line represent the variations of the two energy thresholds. (Bottom panel) Antineutrino energy range. Line styles and legends follow the electron case.}
	\label{fig:Erange}
\end{figure*}

The schematic diagram illustrating the selection of the energy ranges for electrons and neutrinos corresponding to the above discussion are shown in Figure \ref{fig:Erange}. It can be observed that both for electrons and neutrinos, the trend of the maximum energy with scattering angle is similar. This is because in neutron beta decay, most of the energy is transferred to the proton, and the maximum energy selection for electrons and neutrinos is related to the proton mass. The differences in kinematics between electrons and neutrinos themselves are relatively small (See Eqs. (\ref{eq:Emaxpw}-\ref{eq:E1,E2})). Such energy distribution behaviors directly influence the differential width angle distribution of vortex neutron decay, which we will discuss in the following sections.

\section{Numerical results and discussions}
\label{sec:4}
\begin{figure*}[htbp]
	\centering
	\includegraphics[width=0.45\textwidth]{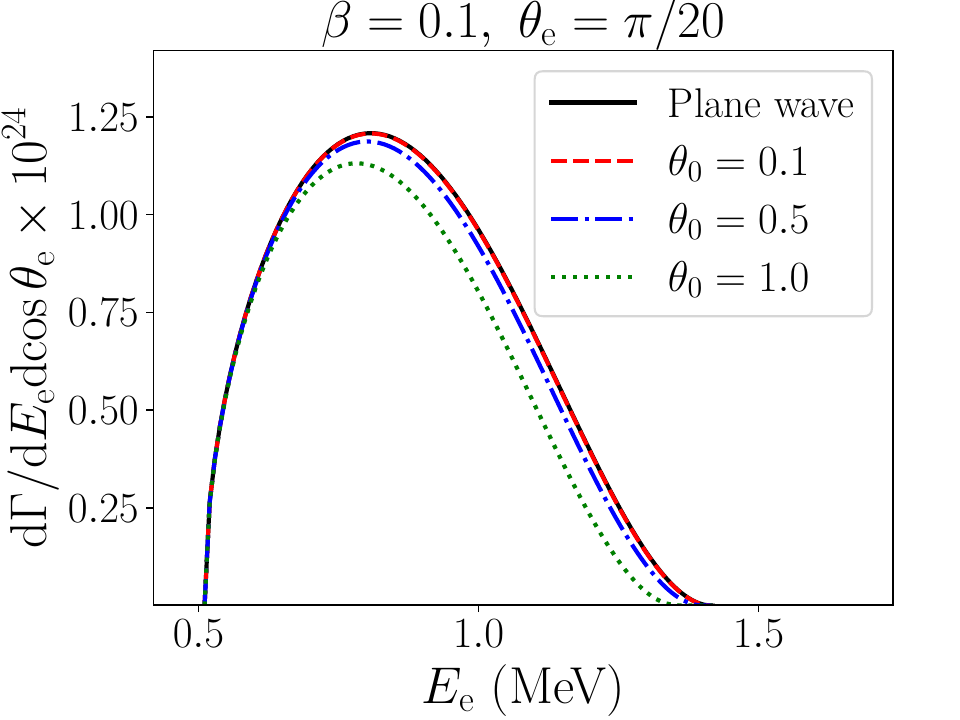}
	\includegraphics[width=0.45\textwidth]{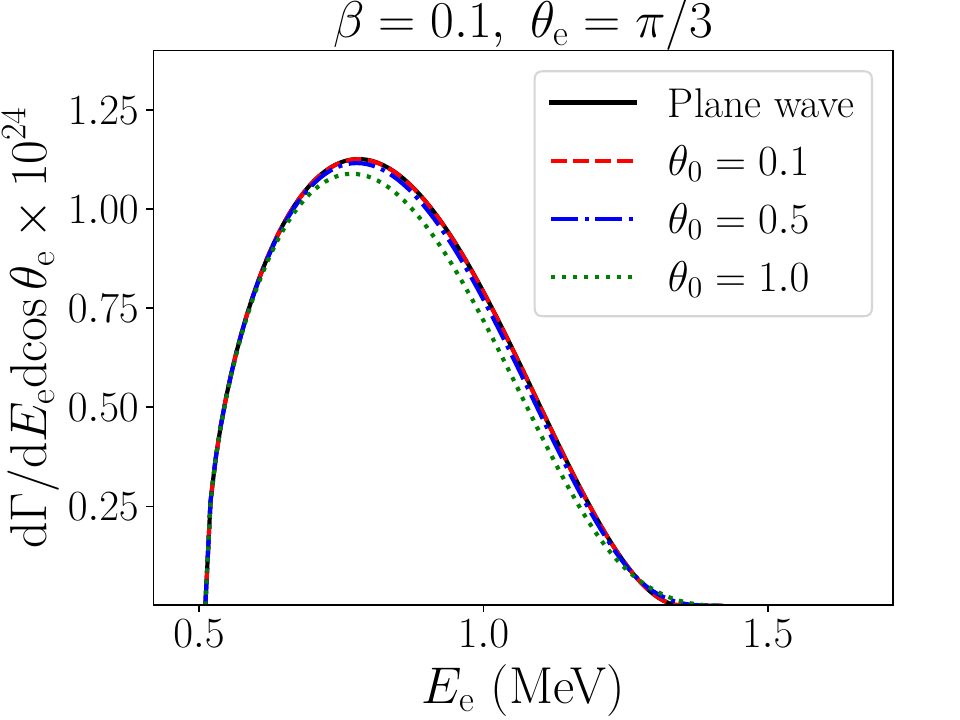}
	\includegraphics[width=0.45\textwidth]{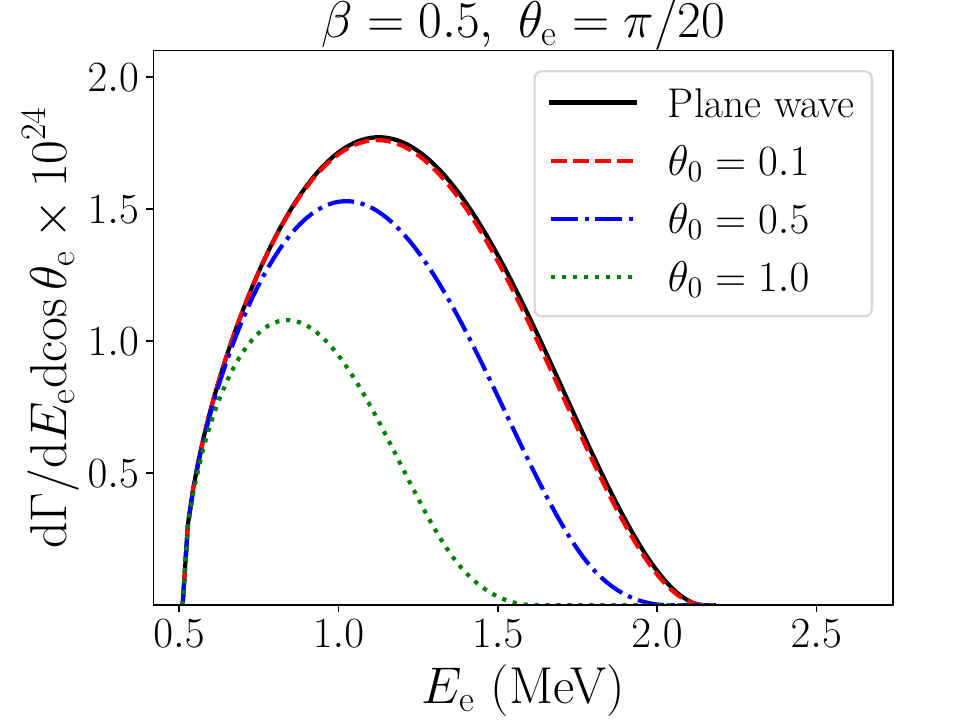}
	\includegraphics[width=0.45\textwidth]{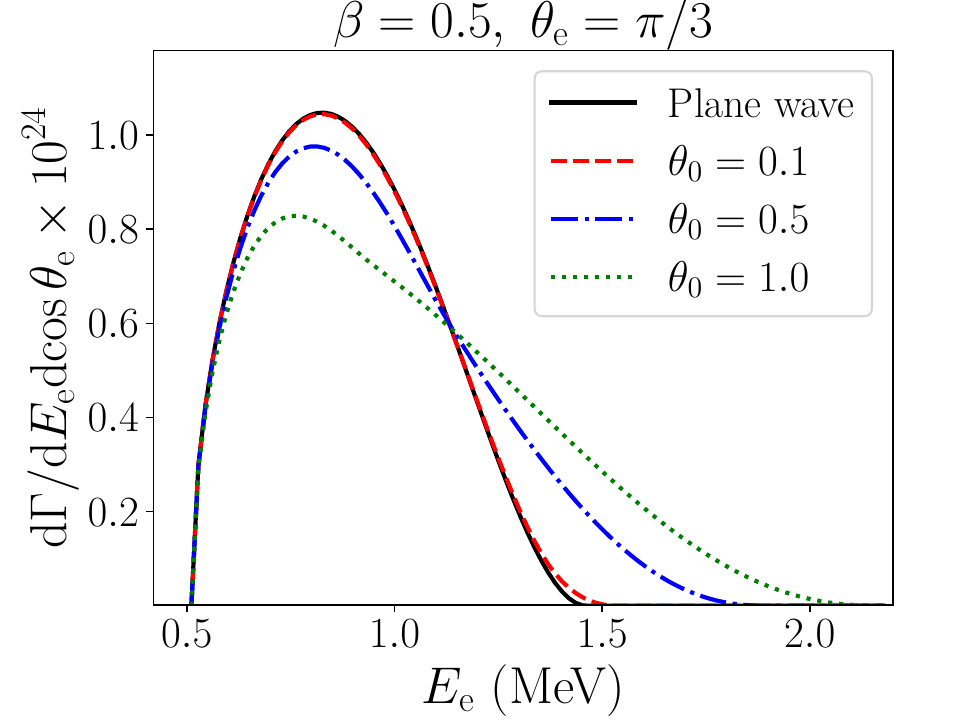}
	\caption{Comparison of the electron energy spectra for plane wave neutrons and vortex neutrons decaying. Assuming neutron speeds of $\beta_n=0.1$ and $0.5$. The electron scattering angles are fixed at $\theta_e=\pi/20$ (left panel) and $\theta_e=\pi/3$ (right panel). The black solid line represents the result for plane wave neutron decay, while the red, blue, and green colored lines represent the vortex neutron decay rate with cone opening angles $\theta_0=$0.1, 0.5, 1.0, respectively.}
	\label{fig:ele-energy}
\end{figure*}

\begin{figure*}[htbp]
	\centering
	\includegraphics[width=0.45\textwidth]{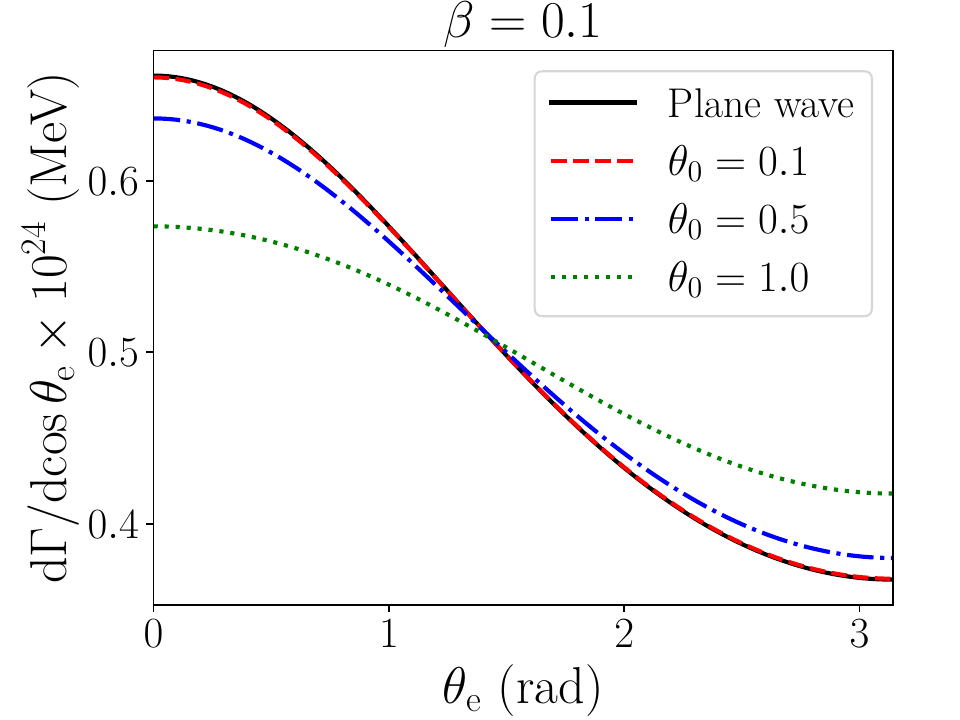}
	\includegraphics[width=0.45\textwidth]{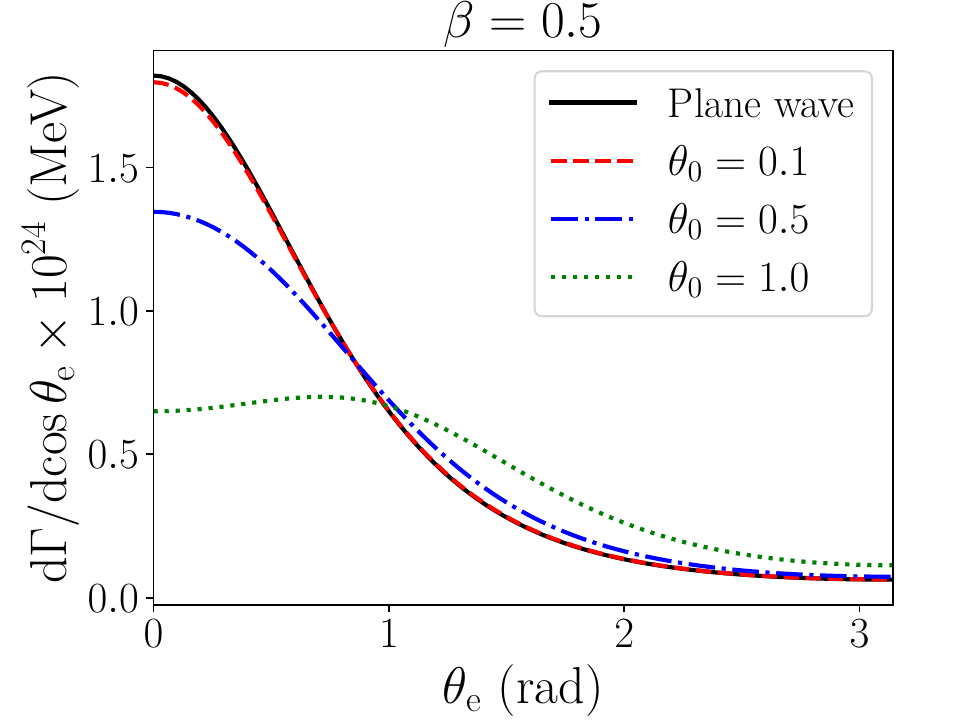}

	\caption{Comparison of the electron scattering angle distributions for plane wave neutrons and vortex neutrons decaying. Assuming neutron speeds of $\beta_n=0.1$ (left panel) and $0.5$ (right panel). The black solid line represents the result for plane wave neutron decay, while the red, blue, and green colored lines represent the vortex neutron decay rate with cone opening angles $\theta_0=$0.1, 0.5, 1.0, respectively.}
	\label{fig:ele-angle}
\end{figure*}

\begin{figure*}[htbp]
	\centering
	\includegraphics[width=0.45\textwidth]{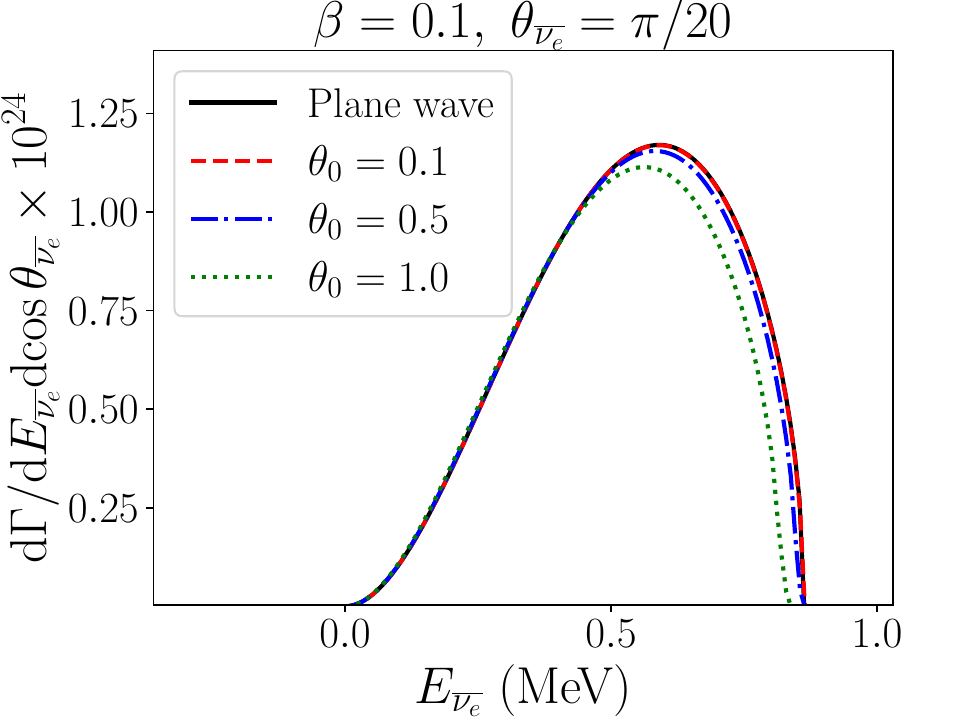}
	\includegraphics[width=0.45\textwidth]{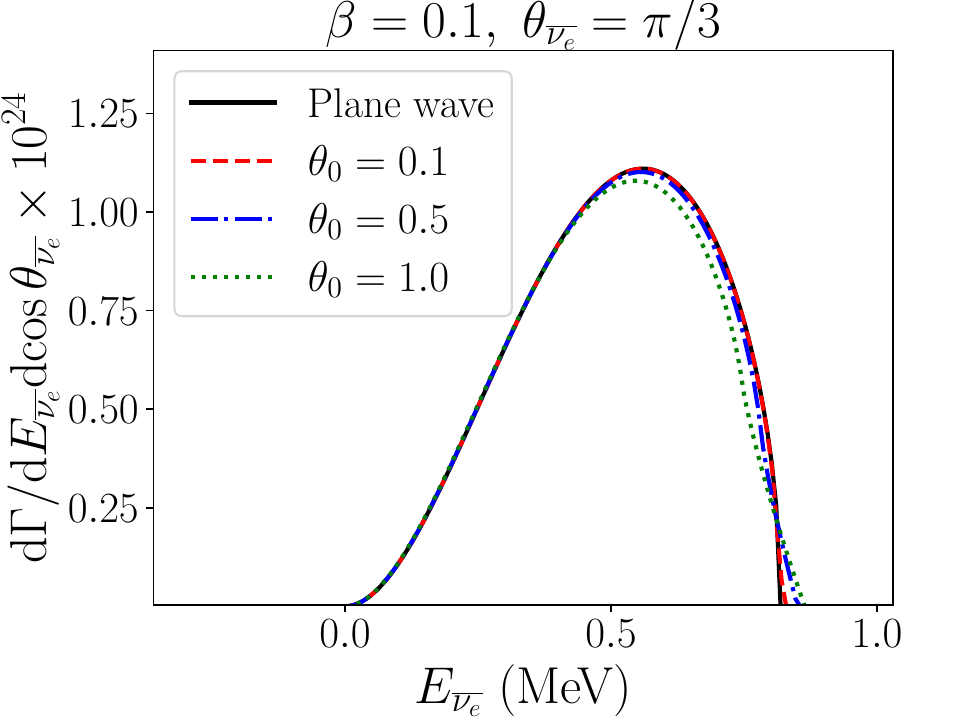}
	\includegraphics[width=0.45\textwidth]{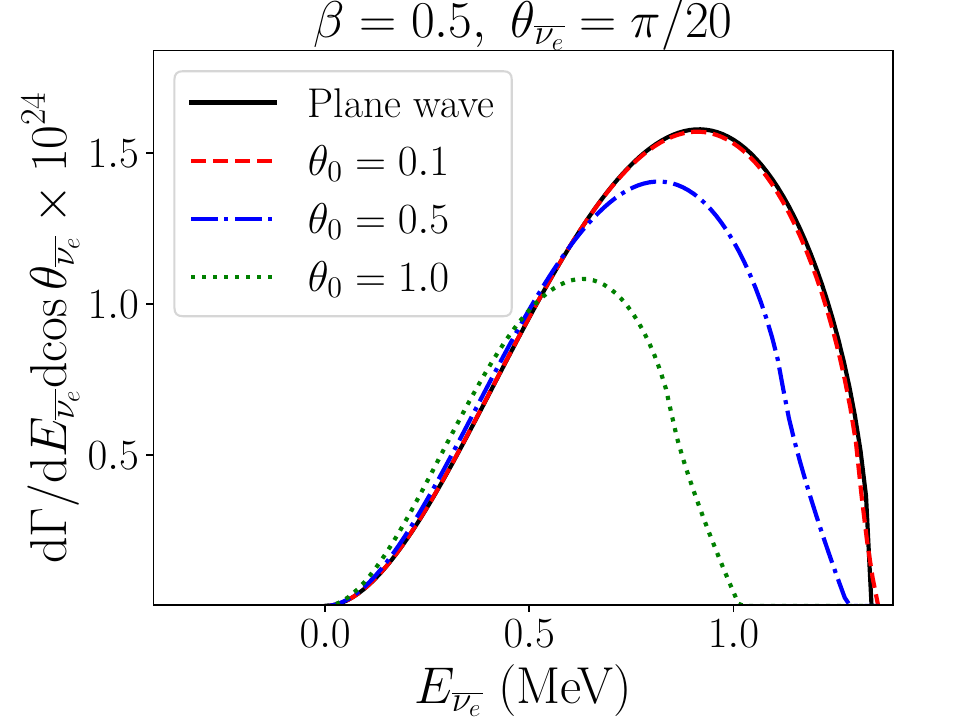}
	\includegraphics[width=0.45\textwidth]{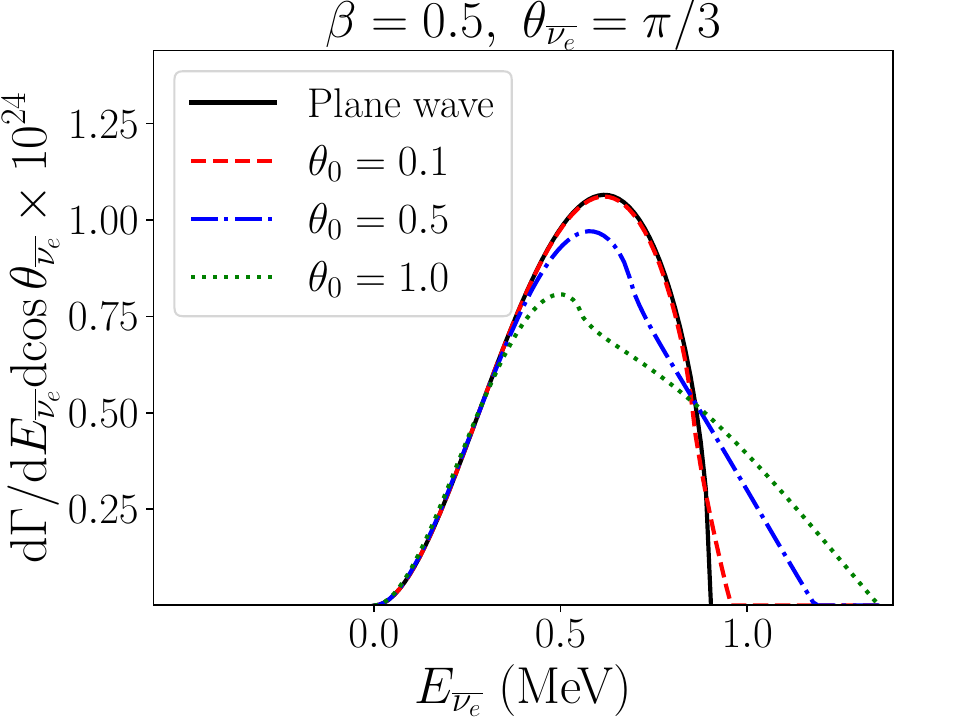}
	\caption{Comparison of the antineutrino energy spectra for plane wave neutrons and vortex neutrons decaying. Assuming neutron speeds of $\beta_n=0.1$ and $0.5$. The antineutrino scattering angles are fixed at $\theta_{\bar{\nu}_e}=\pi/20$ (left panel) and $\theta_{\bar{\nu}_e}=\pi/3$ (right panel). The black solid line represents the result for plane wave neutron decay, while the red, blue, and green colored lines represent the vortex neutron decay rate with cone opening angles $\theta_0=$0.1, 0.5, 1.0, respectively.}
	\label{fig:ne-energy}
\end{figure*}

\begin{figure*}[htbp]
	\centering
	\includegraphics[width=0.45\textwidth]{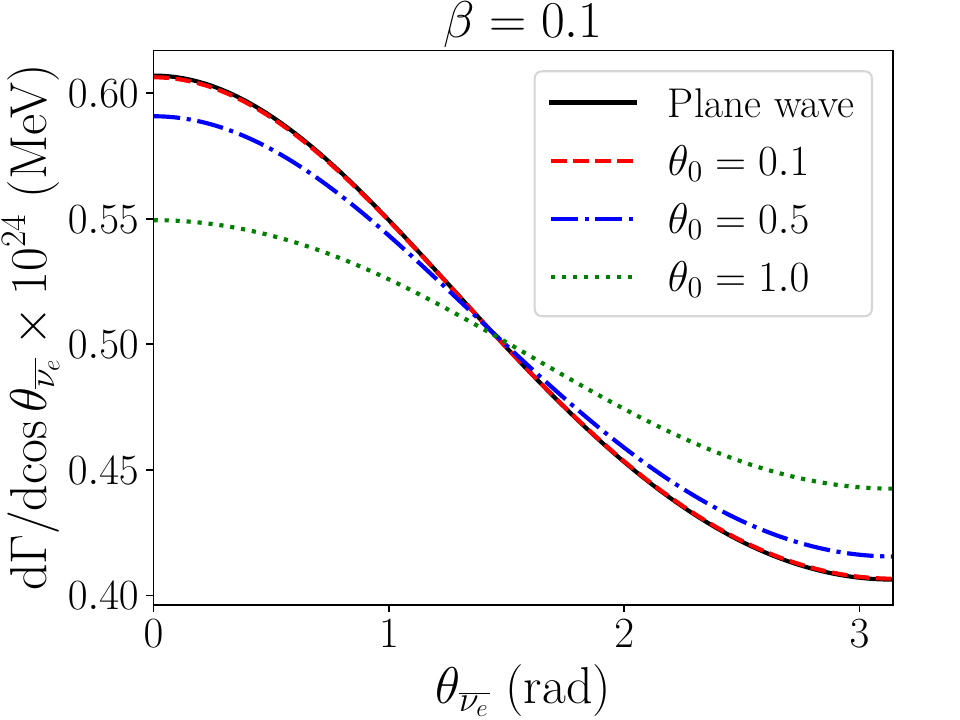}
	\includegraphics[width=0.45\textwidth]{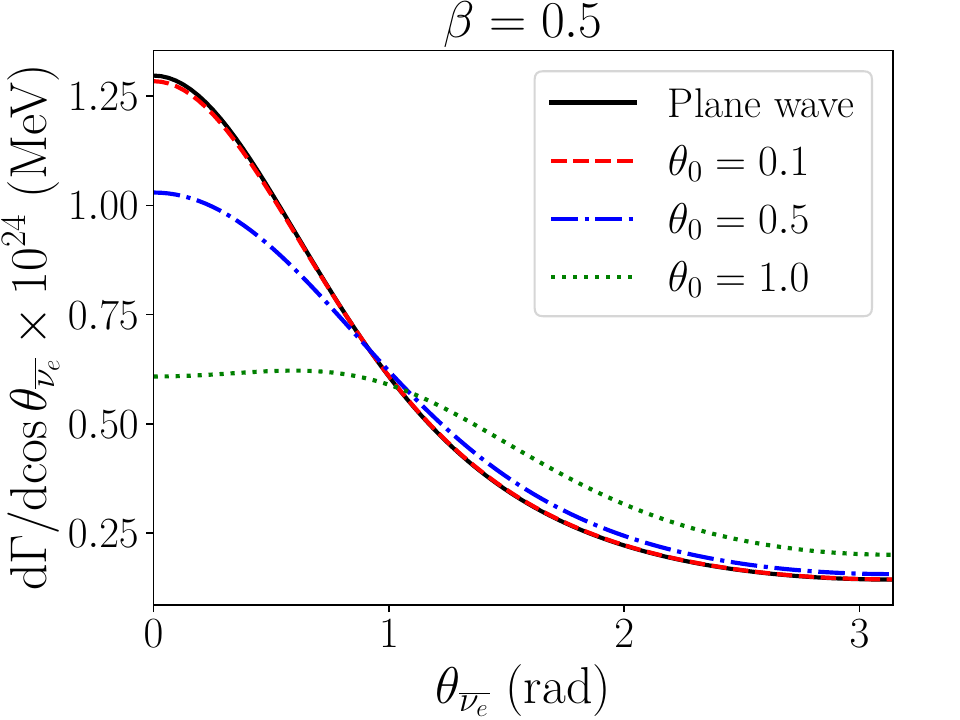}
	
	\caption{Comparison of the antineutrino scattering angle distributions for plane wave neutrons and vortex neutrons decaying. Assuming neutron speeds of $\beta_n=0.1$ (left panel) and $0.5$ (right panel). The black solid line represents the result for plane wave neutron decay, while the red, blue, and green colored lines represent the vortex neutron decay rate with cone opening angles $\theta_0=$0.1, 0.5, 1.0, respectively.}
	\label{fig:ne-angle}
\end{figure*}

In Sec. \ref{subsec: energy}, we discussed the impact of introducing the vortex state on the neutron plane wave distribution on the energy distribution of the decayed particles, implying that different electron energy ranges correspond to different vortex state momentum azimuth angles. The integration for solving the azimuth angle should take into account the influence of different electron energy thresholds. By setting the energy parameters of the moving vortex neutron state as $(p\cdot k)=E_n(E_e-\beta_n\sqrt{E_e^2-m_e^2}\vec{n}_p\cdot\vec{n}_e)$ and plugging it into the Eqs. (\ref{eq:fullwidth}) for calculating the differential width (Considering the integral in Eq. (\ref{eq:inte phip})), we can obtain the electron energy-angle spectrum of the vortex neutron beta decay. Subsequently, integrating over the electron energy yields the differential width dependent only on the electron scattering angle. We assume neutron speeds of $\beta_n=0.1$ and $0.5$ (this is just an example, as controlled high-energy neutron beams are not currently available), with fixed electron scattering angles of $\theta_e=\pi/20$ and $\pi/3$. Then, by adjusting the cone opening angle of the vortex state to $\theta_0=0.1$, $0.5$, and $1.0$, we present the energy spectrum of the vortex neutron decay at fixed electron scattering angles, as shown in Figure \ref{fig:ele-energy}. Subsequently, the integrated differential width angle distribution after integrating out the electron energy is shown in Figure \ref{fig:ele-angle}. We also present the neutrino spectral angle distributions in Figures \ref{fig:ne-energy} and \ref{fig:ne-angle}. In fact, measuring final state neutrinos remains a challenge, but we can still estimate their differential width distribution. To calculate the differential width, one only needs to make some changes to the phase space of the final state, integrating out the momenta of the two other particles except for the neutrino momentum.


An obvious conclusion that can be made is that the beta decay spectral angle distribution of the vortex neutron does not depend on the intrinsic OAM of the initial neutron state. This is because the initial state is a single vortex state and there is no interference, the OAM information is manifested in the phase factor of the vortex state wave function and is canceled out when calculating the scattering matrix element. Therefore, the decay process of a single vortex state particle does not depend on the intrinsic OAM quantum number of the parent particle. However, due to the non-negligible transverse momentum of the vortex state and its unique conical momentum distribution, we can still identify differences from the kinematic characteristics of the final state of vortex neutron decay compared to the results calculated in the plane wave framework.

The main difference between the vortex state and the plane wave state is that the vortex state carries fixed transverse momentum information, which is also reflected in the conical opening angle $\theta_0$ of the momentum distribution. Consequently, significant differences can be observed in both the final state electron energy distribution and the scattering angle distribution as $\theta_0$ varies. Initially, considering the speed $\beta_n$ of the initial neutron, when the conical opening angle $\theta_0$ is fixed, the larger the speed of the neutron (or the modulus of momentum), the greater the distortion in the electron energy spectrum of vortex neutron decay compared to the plane wave framework. This also indicates that appropriately increasing the neutron's energy makes it easier to identify the characteristic differences between plane wave and vortex state decay. Furthermore, different settings of $\theta_0$ can reveal the distinguishability of the spectral angle distribution of vortex neutron decay. A larger $\theta_0$ results in greater differences in the neutron decay spectral angle distribution calculated in the two frameworks. As $\theta_0\to0$, the differences can be neglected, and the decay of the vortex state can revert to the results of the plane wave framework. Additionally, this is also evident in the electron scattering angle distribution (where the electron energy is fully integrated out).

Let us now consider how the relationship between the electron scattering angle $\theta_e$ and the vortex conical opening angle $\theta_0$ affects the decay process of the vortex neutron. In principle, this is related to the two energy thresholds $E_{\mathrm{max},1}$ and $E_{\mathrm{max},2}$ mentioned in the previous section. Vortex neutron decay exhibits unique kinematic characteristics, as shown in Figure \ref{fig:Erange}, where differences in the maximum outgoing electron energy between the vortex and plane wave cases depend on the neutron energy $E_n(\beta_n)$ and the conical opening angle $\theta_0$. When the electron scattering angle satisfies $\theta_e>\theta_0$, meaning the electron is detected outside the vortex cone, the endpoint of the electron spectrum will extend beyond the maximum electron energy corresponding to the plane wave case. In this scenario, according to Eqs. (\ref{eq:Emaxpw}) and (\ref{eq:E1,E2}), $E_{e,\mathrm{max},PW}$ lies between $E_{\mathrm{max},1}$ and $E_{\mathrm{max},2}$. When $\theta_e<\theta_0$, the electron is detected within the cone angle, and the spectral distribution will shift towards the direction of electron energy effectiveness, thus forming a clear contrast with the plane wave neutron decay spectrum. The analysis of the neutrino spectral angle distribution can be analogized to the electron case discussed above, and will not be reiterated here. It is important to emphasize that measuring the neutrino spectrum is challenging under current experimental conditions, but due to the fact that neutrinos are massless within the framework of the Standard Model, the calculated differential widths are more concise in form.

Compared to energy spectrum distributions, the angular distribution of scattering angles can also reflect the differences between plane wave and vortex neutron decay, which similarly depend on the neutron's velocity and the vortex conical opening angle. In conclusion, it can be seen from the above discussion that there are significant differences between vortex state neutron decay and plane wave neutron decay, which can be easily observed from the spectral angle distribution. Furthermore, appropriately adjusting the vortex opening angle and the neutron velocity can amplify the differences between the vortex and plane wave scenarios, which is advantageous for experimental observations.

\section{Conclusion and outlook}
\label{sec:conclusion}
In summary, for the first time we analyze and compute the beta decay process of neutrons in a vortex state, presenting the energy spectra and scattering angle distributions of electrons and neutrinos. We find that the decay of vortex state neutrons does not depend on the neutron's intrinsic OAM, meaning that the calculation process for the decay width of a single vortex state decay will erase the OAM information. Fortunately, due to the presence of transverse momentum information in the vortex state, we can still identify and quantitatively analyze the differences between the decay of vortex state neutrons and plane wave neutrons from the spectral angle distribution influenced by the conical momentum distribution of vortex neutrons. Progress has been made in the preparation of vortex neutrons \cite{Clark:2015rcq,sarenac2018methods,sarenac2019generation,sarenac2016holography,sarenac2022experimental,geerits2023phase}, and research on the scattering processes of vortex neutrons has also advanced significantly \cite{Afanasev:2019rlo,Afanasev:2021uth}. Therefore, studying the decay of vortex state neutrons is timely. The kinematic analysis of three-body decays involving single vortex states, such as beta decay and muon decay \cite{Zhao:2021joa}, is similar, with differences primarily arising from discrepancies in the plane wave scenarios. The decay rate of vortex state particles compared to plane wave state particles is mainly determined by the presence of vortex information in the initial state wave function. In contrast, the scattering process of two-body vortex state particles depends on the intrinsic OAM information of the vortex state due to interference between the vortex state wave functions, as detailed in recent discussions \cite{Zhao:2023cwd}. 

   The realization of high-energy vortex neutrons is of great significance. Currently, there are many researchers exploring methods for generating vortex neutrons \cite{Clark:2015rcq,sarenac2018methods,sarenac2019generation,sarenac2016holography,sarenac2022experimental,geerits2023phase}, although it is still in the early stages, it holds promise. Specifically, there are currently experiments involving neutron or nuclear beta decay to measure the correlation coefficients in the angular distribution \cite{Ivanov:2017mnz,Ivanov:2019bqr,Seng:2024itx}. Neutron states with additional OAM are expected to influence this observable. Although the model we currently employ considers only the single vortex state of the neutron, it can be reasonably anticipated that if the initial-state neutron exhibits interference from vortex states, it would impact the corresponding decay angular distribution correlation coefficients. Additionally, one could envision future measurements of the angular correlation coefficients of relativistically-boosted radioactive nuclei (such as those prepared in FRIB \cite{Johnson:2019sps} and similar facilities). If vortex state neutrons dependent on OAM exist within the relativistically-boosted radioactive nuclei, differences in the angular correlation coefficients may be observed in the measurement results. Finally, the physics processes related to vortex neutrons will provide an opportunity for reexamining atomic nucleus structure and high-energy nuclear reactions.

\section*{Acknowledgments}
This work has been supported by the Strategic Priority Research Program of Chinese Academy of Sciences (Grant NO. XDB34030301) and Guangdong Major Project of Basic and Applied Basic Research (Grant NO. 2020B0301030008).


\bibliography{refs}

\end{document}